\documentclass[superscriptaddress,aps,preprint,pre,nofootinbib]{revtex4}
\usepackage{graphicx}

\long\def\symbolfootnote[#1]#2{\begingroup%
\def\thefootnote{\fnsymbol{footnote}}\footnote[#1]{#2}\endgroup}

\begin{document}

\title{Symmetry, shape and order}

\author{Antonio Trovato}
\affiliation{Dipartimento di Fisica `G. Galilei', Universit\`a di
Padova, Via Marzolo 8, I-35131 Padova, Italy} \affiliation{CNISM,
Unit\`a di Padova, Via Marzolo 8, I-35131 Padova, Italy}

\author{Trinh X. Hoang}
\affiliation{Institute of Physics and Electronics, Vietnamese
Academy of Science and Technology, 10 Dao Tan, Hanoi, Vietnam}
\affiliation{Department of Physics, 104 Davey Laboratory, The
Pennsylvania State University, University Park, Pennsylvania 16802}

\author{Jayanth R. Banavar}
\affiliation{Department of Physics, 104 Davey Laboratory, The
Pennsylvania State University, University Park, Pennsylvania 16802}

\author{Amos Maritan}
\affiliation{Dipartimento di Fisica `G. Galilei',
Universit\`a di Padova, Via Marzolo 8, I-35131 Padova, Italy}
\affiliation{CNISM, Unit\`a di Padova, Via Marzolo 8, I-35131 Padova, Italy}
\affiliation{Sezione INFN, Universit\`a di Padova, I-35131 Padova, Italy}

\begin{abstract}
Packing problems have been of great interest in many diverse contexts
for many centuries. The optimal packing of identical objects has been
often invoked to understand the nature of low temperature phases of
matter. In celebrated work, Kepler conjectured that the densest
packing of spheres is realized by stacking variants of the
face-centered cubic lattice and has a packing fraction of
$\pi/(3\sqrt{2}) \sim 0.7405$. Much more recently, an unusually high
density packing of approximately $0.770732$ was achieved for congruent
ellipsoids. Such studies are relevant for understanding the structure
of crystals, glasses, the storage and jamming of granular materials,
ceramics, and the assembly of viral capsid structures. Here we carry
out analytical studies of the stacking of close-packed planar layers
of systems made up of truncated cones possessing uniaxial symmetry. We
present examples of high density packing whose order is characterized
by a {\em broken symmetry} arising from the shape of the constituent
objects. We find a biaxial arrangement of solid cones with a packing
fraction of $\pi/4$. For truncated cones, there are two distinct
regimes, characterized by different packing arrangements, depending on
the ratio $c$ of the base radii of the truncated cones with a
transition at $c^*=\sqrt{2}-1$.
\end{abstract}

\maketitle

\section{Introduction}

Many natural forms \cite{Thompson}, such as the DNA double helix
\cite{Pieranski,Stasiak} and seed arrays on sunflowers \cite{Klar},
arise from simple geometrical principles rather than from complex
interactions. Symmetry considerations often play a key role in
determining the nature of order of a system
\cite{Chaikin,DiVincenzo}. The simplest model of matter, a collection
of isotropic objects (spheres) exhibits both the isotropic fluid and
the crystalline phases with a phase transition between them on varying
the packing fraction or density \cite{Szpiro}. Dense packing can
result from either the maximization of the packing fraction
\cite{Szpiro} or from the minimization of the area exposed
\cite{Chandler,Stasiak07} to probe objects such as water molecules. Efficient
packing of a system of rods \cite{Onsager} leads again to an isotropic
fluid phase or to uniaxial order with the orientation of the axis of
one of the rods dictating the preferred alignment of nearby
rods. Liquid crystalline phases \cite{Chaikin} exist between a liquid
with no translational order and a crystal with translational order in
all three directions. This is accomplished by employing constituent
particles which are anisotropic -- the molecules of liquid crystals
are not spherical and can form phases with translational order in
fewer than three dimensions and/or orientational order. Recent work
has shown that an unusually high density packing of approximately
$0.770732$ is achieved for congruent ellipsoids
\cite{StillPRL}. Packing studies are relevant for understanding the
structure of crystals, glasses, the storage and jamming of granular
materials and ceramics \cite{j1,j2,j3,j4,j5,j6,j7}.

Here we study the packing of truncated cones and show how the nature
of order of a system composed of identical objects can depend not only
on the symmetry of the object but also on its shape. The geometry of
small self-assembled clusters of cones has been studied in the context
of the geometry of viral capsids \cite{Glotzer_PNAS2007}. The packing
of cones has also been shown to be relevant for understanding the
geometry of amphiphile nanoparticles having a hydrophobic tail and a
hydrophilic head \cite{Tsonchev}. Tightly packed hierarchical
arrangements are obtained by first aggregating the cones into spheres
or cylinders and then packing these in a face-centered-cubic lattice
or a hexagonal Abrikosov flux lattice \cite{Chaikin} respectively. We
will instead first arrange cones within close-packed planar layers and
then stack different layers on top of each other. We begin with a
discussion of the packing of solid cones and then study the more
general case of the dense packing of solid truncated cones.

\section{Results and discussion}

\subsection{Cone packing}

There are three natural close-packed geometries (Figure 1) of a pair
of cones.  The space-filling arrangements of the cones shown in
Figs. 1(d) and 1(e) are effectively two dimensional with biaxial order
arising from the breaking of the uniaxial symmetry of the cone in
order to achieve close packing \symbolfootnote[1]{The arrangement of
discrete cylinders stacked in an Abrikosov lattice is not isotropic in
the plane perpendicular to the cylinder axis -- rather, it is
invariant under discrete rotations (by integer multiples of $\pi/3$)
with three equivalent directions. We will, anyhow, denote as uniaxial
arrangements possessing at least two equivalent directions within the
plane perpendicular to the symmetry axis. Biaxial order is one in
which a privileged direction exists in that plane so that the only
residual symmetry is the invariance under rotations by $\pi$ \cite{Chaikin}. This
broken symmetry arises from the shape of the constituent objects, even
though they are uniaxial.}. A collection of bicones (Figs. 1(d) and
1(e)) or hour glasses (Fig.  1(d)) can also exhibit biaxial
order. There are other high density arrangements of cones obtained by
first assembling them into basic units and densely packing these
units. Figs. 1(f) and 1(g) depict a helical assembly. Fig. 2 shows a
stack of planes with cones arranged as in Figs. 1(d).

We will assume that the cones have flat bases with the opening angle
$\alpha$ at their apex and the slant height $L$. In micelles
\cite{Tsonchev}, the solvent induces a tip-to-tip attraction between
the cones. Here we consider the role of base-to-base stacking in
facilitating planar packing. The cone volume is given by $V_{cone} =
\pi L^3 \sin\alpha \sin\left(\frac{\alpha}{2}\right)/6$. For the
stack of planes shown in Fig. 2, the volume of a
elementary cell is equal to
\begin{equation}
V_{cell}^{pl} = h \cdot d \cdot s = \frac{2}{3} L^3 \sin(\alpha)
\sin\left(\frac{\alpha}{2}\right),
\end{equation}
where $h$, $d$ and $s$ are the dimensions of the cell as described in
Fig. 2. Thus, the packing fraction of the cones in this
case is equal to
\begin{equation}
F = \frac{V_{cone}}{V_{cell}^{pl}} = \frac{\pi}{4},
\end{equation}
which is independent of $\alpha$.

Interestingly, there is an infinite degeneracy in the close-packing
arrangements of cones stacked in planar layers as in Fig. 2(b), due
to the possibility of choosing between positive and negative shifts
each time a new layer is added to the stack. This is
reminiscent of the two-fold stacking choice made at each hexagonal
layer in the random hexagonal close packed structure of spheres
leading to the stacking variants of the face-centered cubic lattice
which share Kepler's optimal packing fraction (0.7405)
\cite{Chaikin}. Remarkably, the packing fraction of identical
cones, $F \sim 0.78539$, is  higher than the latter, in accord  with
the suggestion that spheres cannot be packed as efficiently as other
convex objects \cite{Glotzer_PNAS2007}. The packing fraction of
cones is also higher than the maximum packing fraction recently
found for dense crystal packing of ellipsoids (0.770732)
\cite{StillPRL}.

It is interesting to compare the packing fraction of this biaxial
arrangement with the common assemblies of amphiphile nanoparticles
\cite{Tsonchev}: the spherical micelles in a face-centered-cubic
lattice arrangement and the hexagonal arrangement of cylindrical
micelles. Note that a definitive proof of the most tightly packed
arrangement is highly non-trivial. For the simpler case of the
packing of spheres, Kepler's conjecture was finally proved only
within the last decade \cite{Szpiro}. For a spherical micelle, as
shown by Tsonchev et al. \cite{Tsonchev}, the number of cones in
each sphere is given by $ N_s = \left[\frac{2\pi}{3\gamma -
\pi}\right]$, where $[x]$ denotes the integer part of $x$ and
$\gamma = \arccos\left(
\frac{\cos(\alpha)}{2\cos^2(\alpha/2)}\right)$ for small enough
$\alpha$ such that the cones (or cone bases) form a hexagonal
arrangement in the sphere surface. This assumption is certainly
valid when $\alpha \leq \pi/3$ ($N_s = 11$ for $\alpha = \pi/3$).
$N_s = 3$ for $\alpha \rightarrow 2\pi/3$ from below and equal to 2
for $\alpha$ larger than $2\pi/3$. The packing fraction of spheres
in a face-centered-cubic lattice arrangement is $\pi/(3\sqrt{2})$
\cite{Chaikin}. One therefore finds that the packing fraction of
cones in spherical micelles is given by
\begin{equation}
F_{sph} = \frac{\pi}{24\sqrt{2}} \sin\left(\alpha\right)
\sin\left(\frac{\alpha}{2}\right)  N_s.
\end{equation}
For a cylindrical micelle, the number of cones in an elementary cell,
defined as a cylinder section of thickness $L \sin(\alpha/2) \sqrt{3}
/ 2$ \cite{Tsonchev}, is $N_c = \left[ \frac{2\pi}{\alpha} \right]$
and the volume of such an elementary cell is $V_{cell}^{cyl} = \pi L^3
\sin(\alpha/2) \sqrt{3}$. The packing fraction of cylinders in a
hexagonal arrangement is $\pi/(2\sqrt{3})$. Thus a hexagonal
arrangement of cylindrical micelles formed by flat-based cones yields
a packing fraction of
\begin{equation}
F_{cyl} = \frac{\pi}{36} \sin(\alpha) \left[\frac{2\pi}{\alpha}\right].
\end{equation}
Fig. 3 shows that the biaxial arrangements lead to a denser packing
than both the hexagonal arrangement of cylindrical micelles
\cite{Tsonchev} and the spherical micelle \cite{Tsonchev} in a
face-centered-cubic lattice arrangement. Note that the use of
cones with curved bases improves the packing fraction of both
cylindrical and spherical micelles \cite{Tsonchev}.

\subsection{Truncated cone packing}

We turn now to the more general case of the packing of truncated
cones (conical frustum) shown in Fig. 4 which provides a
natural bridge between a cylinder ($a=b$) and a cone ($b=0$).

Following our previous analysis, we arrange the objects in a
close-packed planar layer, as in Fig. 5, and then stack consecutive
layers on top of each other. A reference frame is attached to each
layer, as in Fig. 5, so that the shift between consecutive layers is
characterized by $(\Delta x,\Delta y,\Delta z)$, the relative
displacement of the two origins. We do not consider rotations (along
the $z$ axis) because they generally lead to a worse packing. Our goal
then is to accomplish close packing by minimizing $\Delta z$ through
the appropriate selection of $\Delta x$ and $\Delta y$. It is crucial
to consider the circles cut out by intersecting the plane $y=0$ with
successive cone layers in the stacking (see Fig. 6). The condition of
mutual tangency of three such circles determines the minimum distance
between successive layers (see Supplementary Material for details).

In general, one find two different degenerate (i.e. yielding the same
minimum $\Delta z$) solutions: one for $r<R$ and the other for
$r>R$. They can be thought of as being related by a mirror symmetry $x
\rightarrow -x; y \rightarrow -y$ applied to the second layer while
keeping the first layer fixed (see Supplementary Material for
details). Note that this choice among two possibilities exists each
time a new layer is added to the stack yielding an infinite degeneracy
in the close-packing arrangements of cones stacked in planar
layers. Upon choosing symmetrically staggered layers, the center
of the circles cut out in the $y=0$ plane arrange themselves on a
regular planar lattice. The symmetry displayed in the latter (and in
the corresponding triangular tiling of the plane $y=0$ obtained by
joining the circle centers) is related to the uniaxiality/biaxiality
of the corresponding packing arrangement.

Remarkably, the optimal stacking and the related symmetries depend on
the value of the ratio $c=b/a$ with a transition point at
$c^*=\sqrt{2}-1$ separating two distinct classes of behavior (see
Methods for a detailed discussion of symmetries and packing fraction
equations in the different regimes). The special case of a cylinder,
$c=1$, characterized by the hexagonal Abrikosov lattice, yields a
tiling of equilateral triangles. In the cylinder-like regime,
$1>c>c^*$, isosceles triangles form a rhombic lattice (see Fig. 7),
whereas in the cone-like regime, $0\le c<c^*$, right angled triangles
result in a rectangular lattice (see Fig. 8).  At the transition point
between the two regimes, $c=c^*$, right angled isosceles triangles
result in a square lattice (see Fig.  9). Cylinders ($c=1$) and
truncated cones at the transition point ($c=c^*$) are special cases
where uniaxiality is maintained since different directions are
equivalent in the plane orthogonal to their axes. In all other cases
($1>c>c^*$ and $0\le c<c^*$) no such symmetry is present, implying
biaxial order.

Fig. 10 shows a plot of the packing fraction $f\left(c\right)$ as a
function of the base radii ratio $c=b/a$ from Equations (\ref{eqf1})
and (\ref{eqf2}). The transition at $c=c^*$ between the cone-like and
the cylinder-like regimes is clearly visible. Most notably, the
packing fraction does not increase monotonically with $c$ in either
regime so that both the transition point ($c=c^*$) and the cone
point ($c=0$) are local maxima for the packing fraction.

\section{Conclusions}

Truncated cones are inherently uniaxial objects. Since they smoothly
interpolate between cylinders and cones we were able to assess the
relevance of shape in dictating the nature of the packing.  The
rigorous determination of the optimal packing is a formidable
problem and the well-packed arrangements we have found can at best
be thought of as conjectures of the best optimal packing. The
packings we have investigated are all based on the simplifying
hypothesis that the optimal solutions are formed from the assembly
(stacking) of close-packed planar layers. Within this assumption, we
have shown that the optimal packing of uniaxial truncated cones is
characterized by  broken symmetry and is in general biaxial with the
exception of the (degenerate) cylindrical case and of a special
value of the 'aspect ratio' $c^*=\sqrt{2}-1$. We have shown that the
tiling of the truncated cone cross-section in the plane orthogonal
to their axes is a useful way to understand the nature of the order,
allowing one to distinguish between two different regimes, $c>c^*$
and $c<c*$. At the transition point separating the two regimes,
truncated cones with $c^*=\sqrt{2}-1$ have interesting symmetry
properties. The packing
fraction that is achieved by truncated cones is remarkably high.\\

We conclude with a speculation pertaining to the building blocks of
protein native state structures -- uniaxial helices and biaxial
sheet. There is a simple way of understanding how helices can emerge
as a natural compact form adopted by a uniaxial tube. There is no
equally simple way of rationalizing the existence of zig-zag strands
which assemble into almost planar, biaxial sheets. Our results above
suggest that even uniaxial objects, because of their shape, can
exhibit broken symmetry and biaxial order. Instead of thinking of the
packing of separate objects, consider now the case of a linear chain
molecule made up of objects tethered together. There is a special axis
at any location along the chain defined by the positions of the
adjoining tethered objects. This leads naturally to the requirement
that the constituent objects at least have uniaxial symmetry rather
than be isotropic objects or spheres. Such a chain of anisotropic
coins can be thought of as having a tube-like geometry in the
continuum limit. The helix is a natural compact arrangement of a
flexible tube and strikingly a tightly wound space-filling helix has
the same pitch-to-radius ratio as $\alpha$-helices in proteins, which
are relatively short polymer chains \cite{Nature00}. Figs. 1(f) and
1(g) depict a helical confirmation \cite{Pauling2} of a short chain of
bicones (Fig. 1(h)). Fig. 1(d) shows an alternate well packed
conformation of the chain. Such biaxial, planar sheet-like structures
\cite{Pauling} are employed by nature not only in the amyloidin
structure \cite{Fowler,Dobson} of proteins but also as a building
block, along with the helix, of protein native state
structures. Furthermore, there is a coordinated positioning of the
amino acid side chains perpendicular to the plane of the sheet. The
simple calculations we have presented here illustrate how, in
principle, both the uniaxial helix and the biaxial sheets can emerge
as the anisotropic building blocks of protein structures.

\section{Methods}

The minimum $\Delta z$ obtained on stacking two planar layers
determines the packing fraction $F = V_{tc} / V_{\rm cell}$ where
$V_{\rm cell} = h \Delta z \left(a+b\right)$ is the volume of the
elementary cell in the periodic arrangement of truncated cones and
$V_{tc} = (\pi h/3) \left(a^2+ab+b^2\right)$ is the volume of the
truncated cone. Different regimes are found depending on the value of
$c$. More details are given in Supplementary Material.

\subsubsection{Cylinder-like regime: $c>c^*$ }

The packing fraction is

\begin{equation}
F = \frac{\pi}{6}\frac{1+c+1/c}{\sqrt{1+2/c}}.
\label{eqf1}
\end{equation}

The triangle obtained on connecting the centers of the three circles
involved in the mutual tangency condition (see Fig. 6) is
isosceles, since two of the circles have the same radius ($r=a$ or
$r=b$). Note that for these solutions the only effective mirror
symmetry is $x \rightarrow -x$, because $\Delta y = h$ or $\Delta y
= 0$.

For the limiting case of a cylinder ($a=b$) we correctly obtain
$\frac{\pi}{2 \sqrt{3}} = 0.9069\dots$ and the triangle defined
above becomes equilateral. The restoration of uniaxiality in this
limit is underscored by the invariance of the resulting triangular
tiling of the $\left(x,z\right)$ plane, under rotation by an integer
multiple of $\pm 60^{o}$.

The breaking of the uniaxial symmetry occurs as soon as $b$ is
strictly smaller than $a$ and it is accompanied by: i) all the
triangles becoming isosceles and ii) only half of them being
associated with the mutual tangency condition (see Fig. 7) yielding
the two-fold degeneracy discussed above. Note that on moving along the
$y$ axis, the circles formed in the $\left(x,z\right)$ plane change
their radii.

\subsubsection{Cone-like regime: $c<c^*$}

The packing fraction is

\begin{equation}
F = \frac{\pi}{6}\frac{\left(3+c\right)\left(1+c+1/c\right)}{\left(1+c\right)\left(1+2/c\right)}.
\label{eqf2}
\end{equation}

Note that for these solutions the only effective mirror symmetry is
$y \rightarrow -y$, because $\Delta x = a + b$ or $\Delta x = 0$
(see Figs. 8 and 11).

In the limiting case of a regular cone ($b=0$) we correctly obtain $F
= \frac{\pi}{4} = 0.7854\dots$.

The triangles defined above are always right angled but never
isosceles. The triangular tiling of the $\left(x,z\right)$ plane
thus results in a rectangular tiling. Again, only half of the
triangles are associated with the mutual tangency condition (see
Fig. 8) and uniaxial symmetry is broken (i.e. no
rotation symmetry is present in the plane).

\subsubsection{Transition point: $c = c^*$}

The special value $c=c^*$, i.e. $b = \left(\sqrt{2} - 1\right)
a$, separates the two above regimes. At $c=c^*$ the packing fraction
is

\begin{equation}
F = \frac{\pi}{6} \left(3 - \sqrt{2}\right) = 0.8303\dots .
\end{equation}

The triangles defined above are isosceles right angled and the
degeneracy disappears. Indeed the two degenerate solutions merge
one into the other and one obtains a square tiling in the
$\left(x,z\right)$ plane (see Fig. 9). Uniaxiality is restored in this
special case because the square tiling is invariant under rotation by
an integer multiple of $\pm 90^{o}$. The triangles are no longer
isosceles and the square tiling become rectangular for any $c<c^*$
(see Fig.  8), whereas the triangles are no longer right angled for
any $c>c^*$ (see Fig. 7).

\acknowledgments

This work was supported by PRIN no. 2005027330 in 2005, INFN,
the NSC of Vietnam, and the Vietnam Education Foundation.

\newpage

\begin{figure}[t]
\includegraphics[height=14.cm]{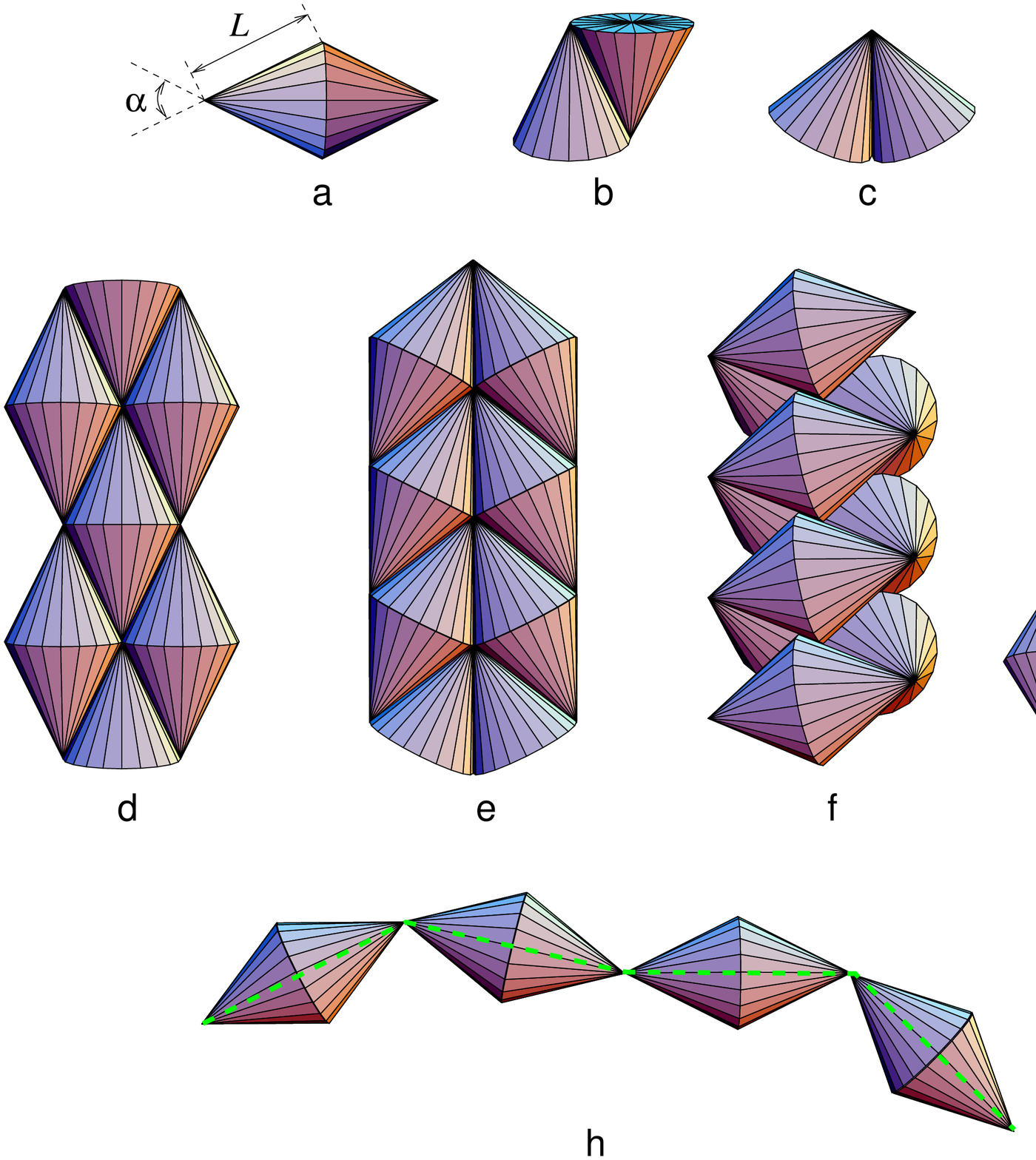}
\caption{Three close-packed configurations of two
identical cones: a) the cones form a bicone, b) the cones point in
opposite directions, and c) the cones axes are not parallel. d) and e)
Two close-packed configurations of many identical cones. Note that
both arrangements are necessarily planar to achieve close packing. f)
and g) Two views of a close-packed helix, a common motif in
biomolecular structure h) A linear chain made of 4 identical
bicones. Configurations (d), (f) and (g) are viable close-packed
arrangements for a chain molecule.} \label{fig1}
\end{figure}
\clearpage

\begin{figure}[t]
\includegraphics[height=16.cm]{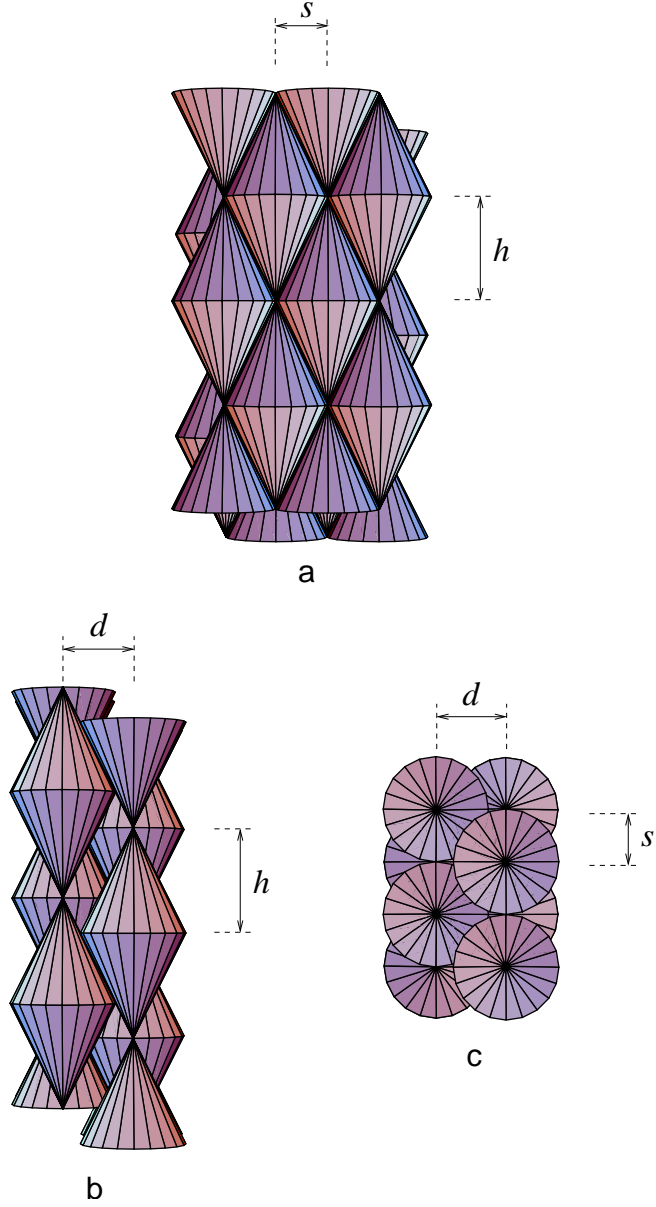}
\caption{Stack of planes formed by cones. a), b)
and c) are three side views of a stack of two planes shown in
Fig. 1(d).  In (b), the vertical shift of the first layer
with respect to the second layer is $L\cos(\frac{\alpha}{2})/3$, where
$L$ is the slant height of the cone and $\alpha$ is opening angle of
the cone at its apex.  The layer separation is $d = \frac{4}{3}
L\sin(\frac{\alpha}{2})$.  The lateral shift of successive planes is
$s = L\sin(\frac{\alpha}{2})$, and the height of the elementary cell
is $h = L\cos\left(\frac{\alpha}{2}\right)$.} \label{stack6}
\end{figure}
\clearpage

\begin{figure}[t]
\includegraphics[height=10.cm]{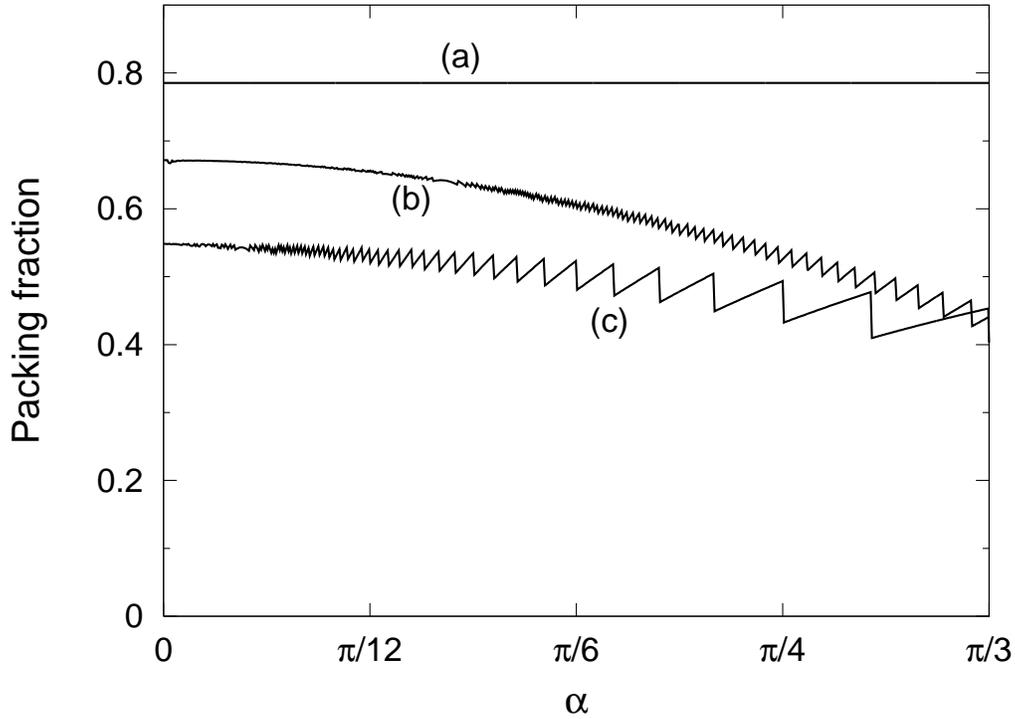}
\caption{Dependence of the packing fraction of
cones with a flat base on the cone's opening angle at the apex,
$\alpha$.  a) The packing fraction of the stack of planes shown in
Fig. 2. This packing fraction is independent of the angle
$\alpha$ and is equal to $\frac{\pi}{4}$. b) Packing fraction of a
face-centered-cubic lattice of spherical micelles \cite{Tsonchev}
formed by the cones. c) Packing fraction of a hexagonal arrangement of
cylindrical micelles \cite{Tsonchev}. One can show that appropriately
stacked planes shown in Fig. 1(e) also yield packing fractions greater
than the micellar arrangements.} \label{pack5}
\end{figure}
\clearpage

\begin{figure}[t]
\includegraphics[height=5.cm]{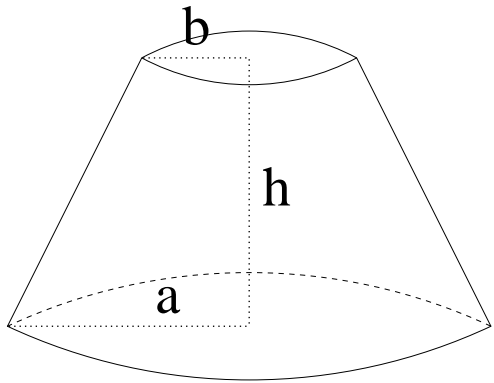}
\caption{Conical frustum of height $h$, base radius
$a$, and top radius $b$ ($a\ge b$).} \label{trunc}
\end{figure}
\clearpage

\begin{figure}[t]
\includegraphics[height=10.cm]{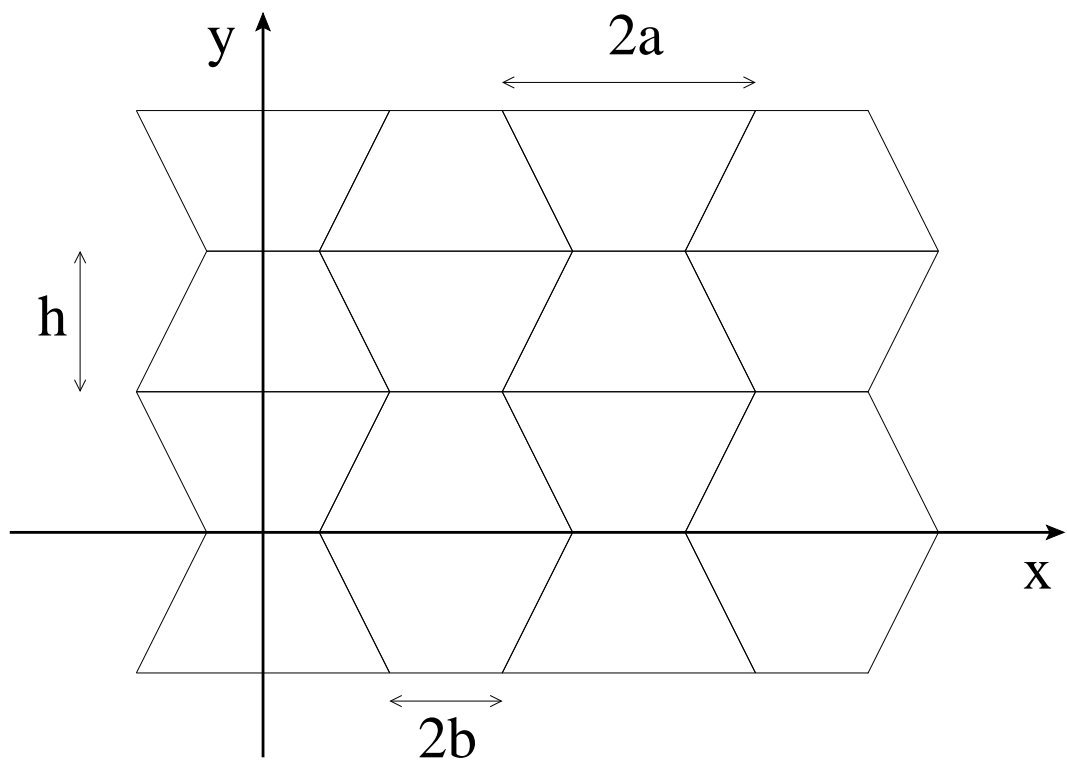}
\caption{Close-packed planar layer of truncated
cones.}
\label{trlayer}
\end{figure}
\clearpage

\begin{figure}[t]
\includegraphics[height=10.cm]{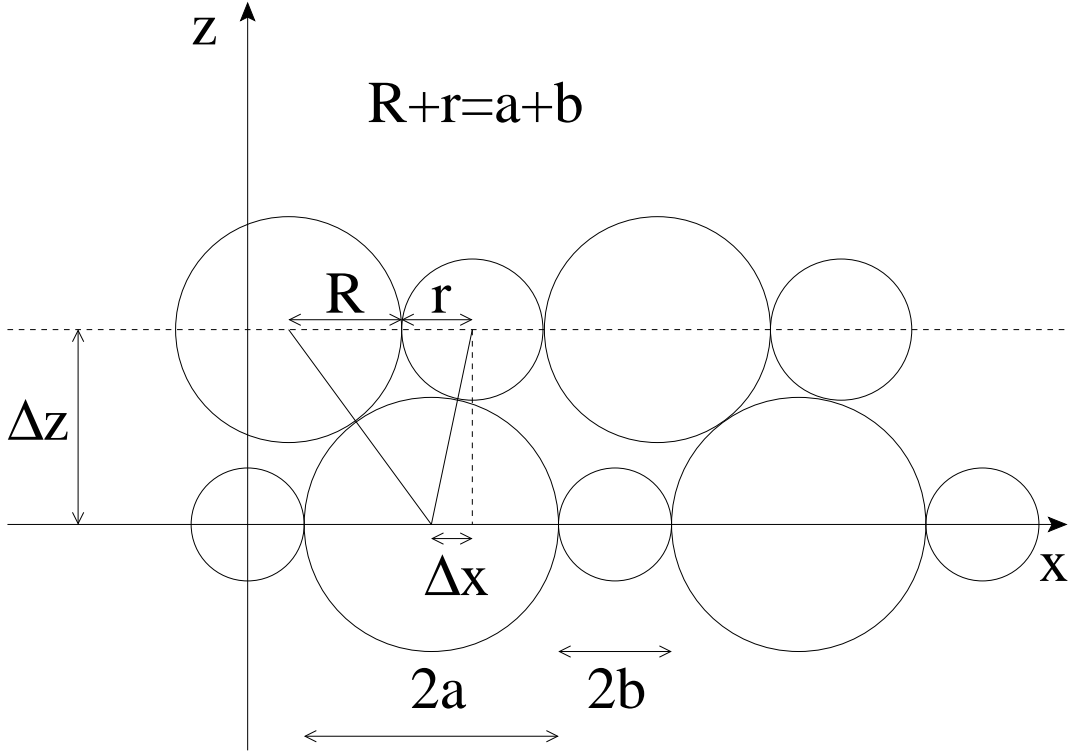}
\caption{Intersection of the plane $y=0$ with two layers of the
stack. The lower circles belong to cones in the first layer ($z=0$)
and are either base or top circles with radius $a$, $b$,
respectively. The upper circles belong to cones in the second layer
($z=\Delta z$). Their radii $r$, $R$ ($b\le r,R \le a$) are determined
by the shift $\Delta y$ along the $y$ axis and always fulfill the
relation $R+r=a+b$. The shifts $\Delta x$ and $\Delta z$ between the
two layers in the $(x,z)$ plane are shown. The condition of mutual
tangency of the `big' base circle from the first layer with both
circles from the second layer is shown.} \label{trcut}
\end{figure}
\clearpage

\begin{figure}[t]
\includegraphics[height=10.cm]{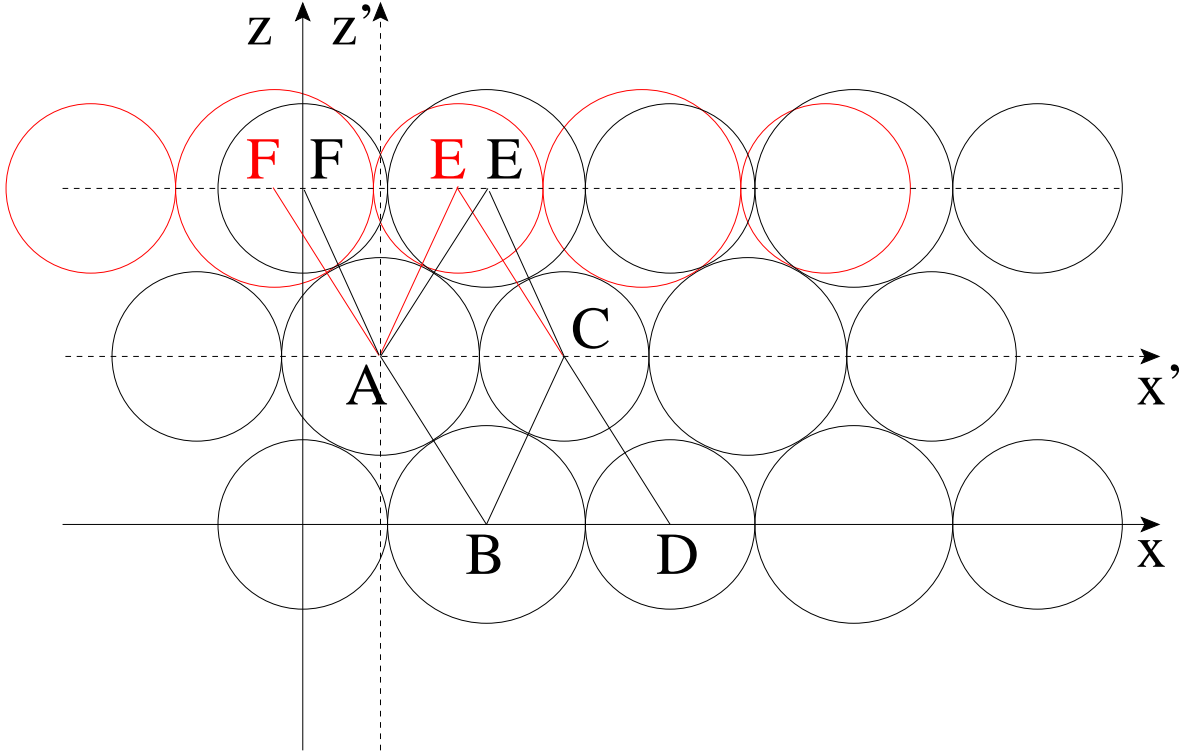}
\caption{Intersection of the plane $y=0$ with multiple layers of the
stack in the cylinder-like regime $c>c^*$. Note that $\Delta x\ne0$
whereas $r=a$ and $R=b$. The two different ways of adding the third
layer to the stack are shown (in black and red). The triangles $ACB$
and $BCD$ are isosceles because
$\overline{AC}=\overline{BD}=\overline{BC}=a+b$ whereas
$\overline{AB}=\overline{CD}=2a$. For triangle $ACB$, all sides pass
through tangency points (``full'' triangle) whereas the same does not
happen for triangle $BCD$ (``empty'' triangle). The choice one has in
placing the third layer is equivalent to selecting how to continue the
isosceles triangular tiling or equivalently where to place the
``empty'' triangles. The triangle $AEC$ is ``full' in the black case
and ``empty'' in the red case. Note that in the reference frame
$\left(x',z'\right)$ defined by the second layer the two possible
placements (black and red) of the third layer are related to each
other by the reflection $x' \rightarrow - x'$. In the cylinder limit,
$c=1$, one recovers equilateral triangular tiling, all triangles are
``full'', the triangle $AEF$ remains the same under the reflection $x'
\rightarrow - x'$ and no degeneracy is present anymore. Rotations by
$\pi$ about the $x$, $y$, or $z$ axis relate different stacking
variants to each other, showing that the packing of truncated cones in
this regime is biaxial \cite{Chaikin}, with special directions $x$,
$y$, $z$.} \label{trsymm}
\end{figure}
\clearpage

\begin{figure}[t]
\includegraphics[height=10.cm]{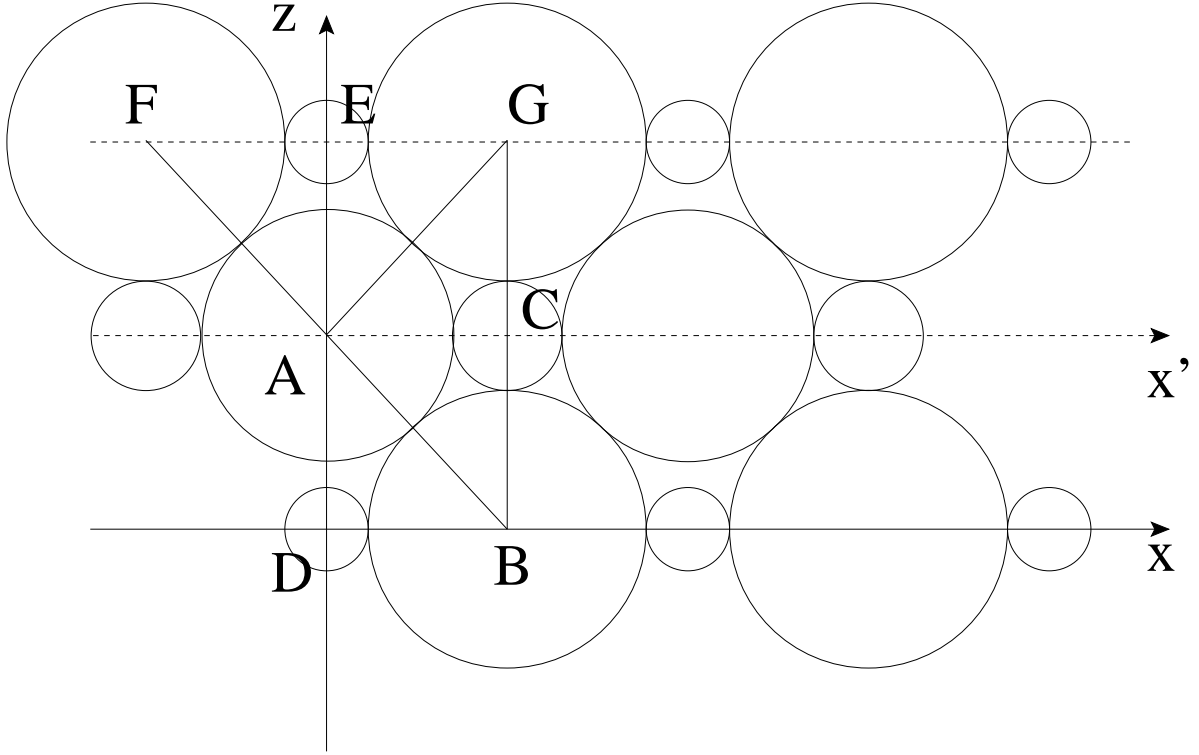}
\caption{Intersection of the plane $y=0$ with multiple layers of the
stack in the cone-like regime $c<c^*$. Note that $\Delta x=0$ and
$r\ne b$, $r\ne a$. The triangles $ACB$ and $ABD$ are right angled but
not isosceles because $\overline{AC}=\overline{BD}=a+b=r+R$,
$\overline{BC}=\overline{AD}=a+r>a+b$, and $\overline{AB}=a+R$. For
triangle $ACB$ all sides pass through tangency points (``full''
triangle) whereas the same does not happen for triangle $ABD$
(``empty'' triangle). The triangle $AEF$ is changed into the
equivalent ``empty'' triangle $AEG$ under the reflection $x'
\rightarrow - x'$. The two degenerate solutions are not visible in the
$\left(x,z\right)$ plane because they are connected by the reflection
$y' \rightarrow - y'$. Rotations by $\pi$ about the $x$, $y$, or $z$
axis relate different stacking variants to each other, showing that
the packing of truncated cones in this regime is biaxial
\cite{Chaikin}, with special directions $x$, $y$, $z$.}
\label{trsymm3}
\end{figure}
\clearpage

\begin{figure}[t]
\includegraphics[height=10.cm]{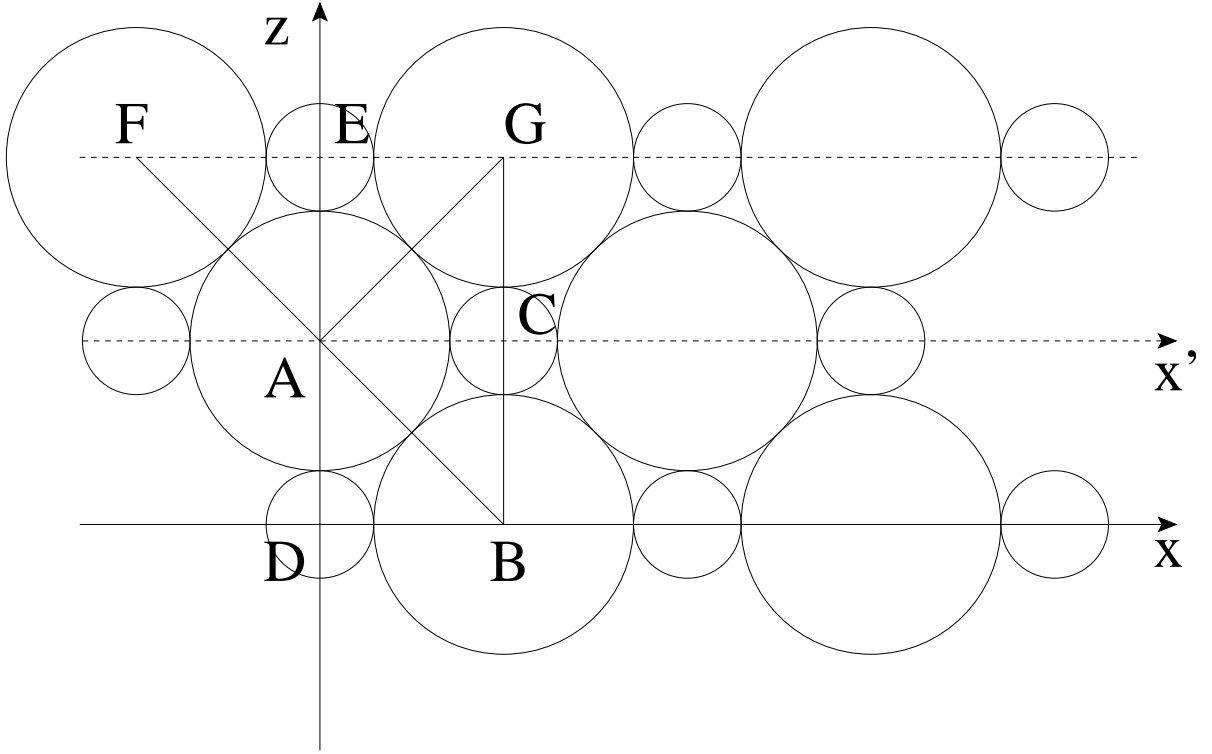}
\caption{Intersection of the plane $y=0$ with
multiple layers of the stack at the transition point $c=c^*$. Note that
$\Delta x=0$ and $r=b$ ($\Delta x=a+b$ and $R=a$ corresponds to the
same solution).  The triangles $ACB$ and $ADB$ are right angled and
isosceles because
$\overline{AC}=\overline{BD}=\overline{BC}=\overline{AD}=a+b$, whereas
$\overline{AB}=2a$. For each triangle all sides pass through tangency
points, i.e. all triangles are ``full''. The triangle $AEF$ is changed
into the equivalent triangle $AEG$ under reflection $x' \rightarrow -
x'$, so that no degeneracy is present in this case.}
\label{trsymm2}
\end{figure}
\clearpage

\begin{figure}[t]
\includegraphics[height=12.cm]{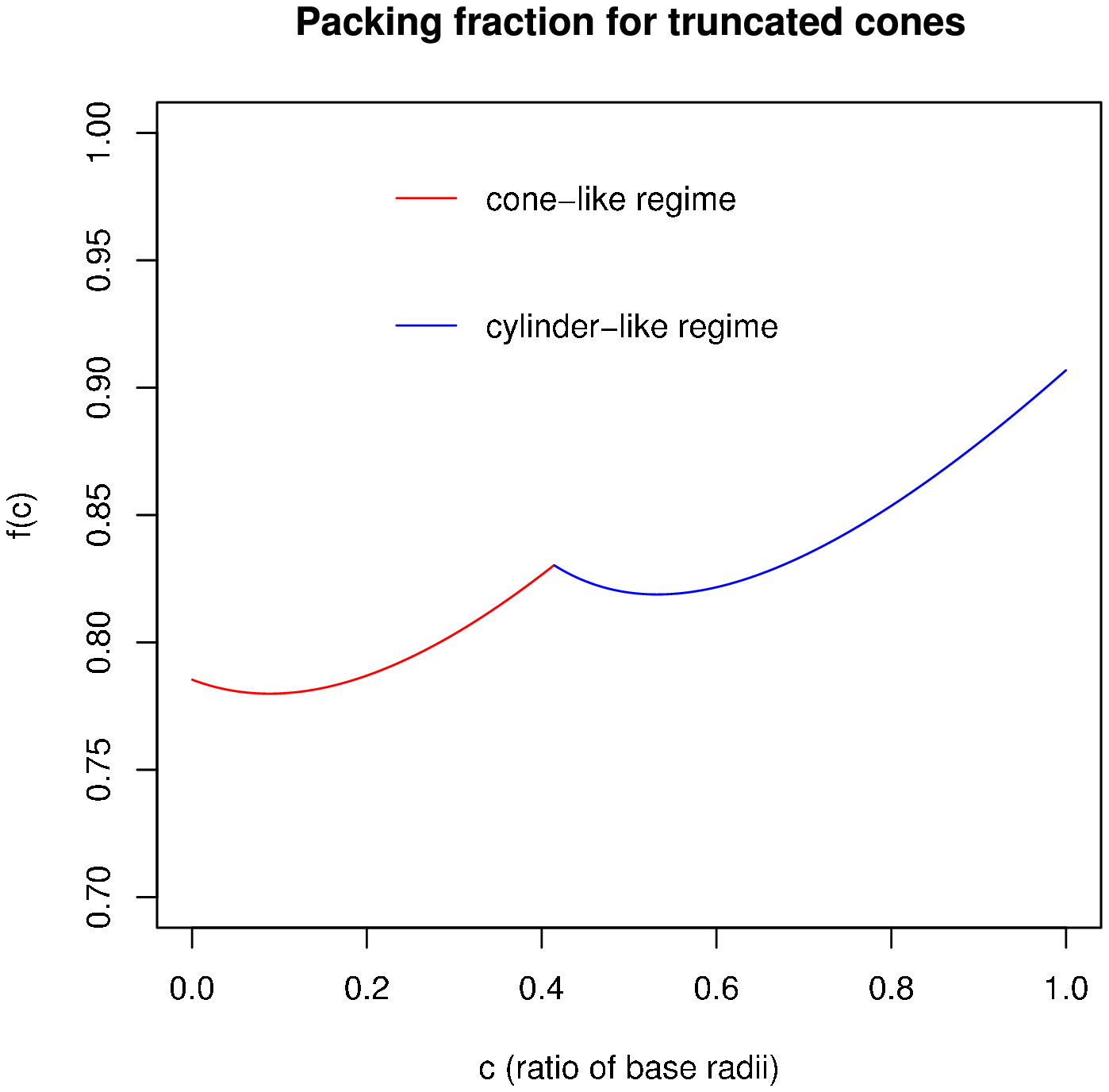}
\caption{Packing fraction $f\left(c\right)$ as a
function of the base radii ratio $c$ from equations (\ref{eqf1})
(red line, cone-like regime) and (\ref{eqf2}) (blue line, cylinder-like
regime).}
\label{packfr}
\end{figure}
\clearpage

\begin{figure}[t]
\includegraphics[height=10.cm]{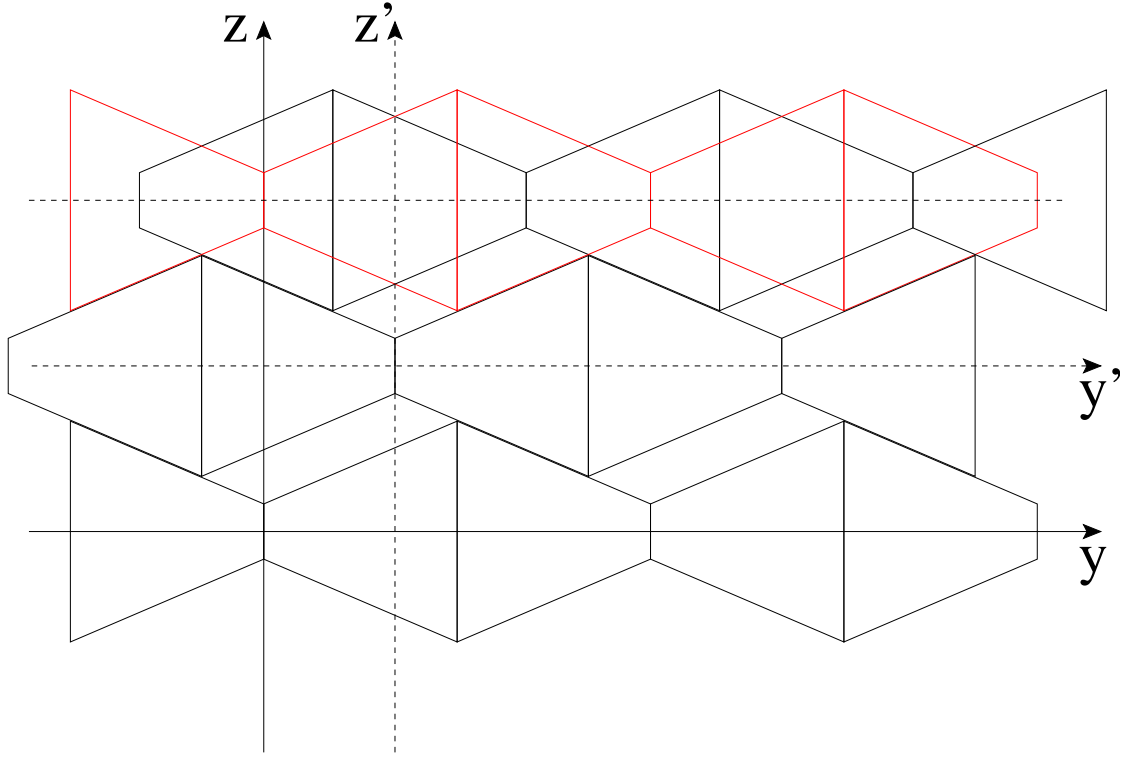}
\caption{ Intersection of the plane $x=0$ with
multiple layers of the stack in the cone-like regime $c<c^*$. Note
that $\Delta y\ne0$. The two different ways of adding the third layer
to the stack are shown (in black and red). Note that in the reference
frame $\left(y',z'\right)$ defined by the second layer, the two
possible placements (black and red) of the third layer are related to
each other by the reflection $y' \rightarrow - y'$. In the
cylinder-like and threshold regime $c\ge c^*$, $\Delta y=0$ if $r=a$,
and no void is observed when the packing is viewed in the
$\left(y,z\right)$ plane.} \label{trsymm4}
\end{figure}
\clearpage

\section{Supplementary text}

\subsection{Packing of truncated cones}

We wish to study the packing of truncated cones (see Fig. 4). We make
the plausible assumption that the best packing can be found by first
arranging the truncated cones in a close-packed planar layer, as in
Fig. 5 and then stacking consecutive layers on top of each other. A
reference frame is attached to each layer, as in Fig. 5, so that the
shift between consecutive layers in the stacking is defined as the
shift $(\Delta x,\Delta y,\Delta z)$ between the origins of the
corresponding frames. Rotations (along the $z$ axis) can be defined
similarly, but we assume that the rotation of a layer with respect to
an adjacent one in the stacking leads to a worse packing.

The results presented in the following are obtained by finding the
values of $\Delta x$ and $\Delta y$ that minimize $\Delta z$
compatibly with steric constraints (in {\it Cone Packing} under {\it
Results and Discussion}, the ``lateral shift'' $s$ is $\Delta x$, the
``vertical shift'' is $\Delta y$, and the layer separation $d$ is
$\Delta z$).

In Fig. 6, the intersection of the plane $y=0$ with two layers of the
stack is plotted for a generic shift $\Delta y$ along the $y$ axis. The
radii $R$, $r$, of the resulting circles cut from the cones in the
second layer are determined by $\Delta y$:

\begin{equation}
R = b + \frac{\Delta y}{h}\left(a - b\right) \qquad
r = a - \frac{\Delta y}{h}\left(a - b\right)
\label{dy}
\end{equation}

For any given $\Delta y$ and $\Delta x$, the minimum $\Delta z$ is
obtained by first imposing the condition of double tangency of the
`big' base circle from the first layer with both circles from the
second layer. The double tangency (already shown to occur in Fig. 6)
implies that $\Delta x$ and $\Delta z$ are determined as a function of
$\Delta y$, or alternatively of $r$ (see Eq. \ref{dy}).

\begin{eqnarray}
\Delta x & = & - a + r \frac{3a + b}{a + b}\;\,\\ \left(\Delta
z\right)^2 & = & 4 r a \left(a + b - r\right)\frac{2a + b}{\left(a +
b\right)^2} \;.
\end{eqnarray}
$\left(\Delta z\right)^2$ is then minimized with
respect to $r$.

Noting that the solutions need to satisfy $0\le\Delta x\le a+b$ and
$b\le r \le a$ (i.e., $0\le \Delta y \le h$), one gets two different
regimes for the optimal stacking depending on the ratio $b/a$. Note
that there is a combined periodicity in the $\left(x,y\right)$ plane
so that $\Delta x=0$, $r=a$ ($\Delta y=0$) is the same as $\Delta
x=a+b$, $r=b$ ($\Delta y=h$).

The optimal stacking depends on the value of the ratio $c=b/a$. There
is a special transition point $c^*=\sqrt{2}-1$ separating two distinct
regimes according to whether $c<c^*$ or $c>c^*$. In general, one finds
two different degenerate (i.e., yielding the same minimum $\Delta z$)
solutions: one for $r<R$ and one for $r>R$. They can be thought to be
related by a mirror symmetry $x \rightarrow -x; y \rightarrow -y$
applied to the second layer while keeping the first layer fixed. This
choice among two possibilities has to be taken each time a new layer
is added to the stack implying an infinite degeneracy in the
close-packing arrangements of cones stacked in planar layers.

\vspace{1.0cm}

\subsubsection{Cylinder-like case}

\vspace{1.0cm}

If $c>c^*$ we are close to the cylinder $c=1$ case.

The two solutions are

\begin{eqnarray}
& & r_1 = b\:;\quad \Delta x_1 = \frac{b^2 + 2ab - a^2}{a + b}\:;\quad
\Delta y_1 = h \\ & & r_2 = a\:;\quad \Delta x_2 = \frac{2a^2}{a +
b}\:;\quad \Delta y_2 = 0
\end{eqnarray}

yielding

\begin{equation}
\Delta z = \frac{2a}{a + b}\sqrt{b\left(2a + b\right)}
\end{equation}

\noindent and a packing fraction ($F=V_{tc}/V_{\rm cell}$, where
$V_{\rm cell}=h\Delta z\left(a+b\right)$ is the volume of the
elementary cell in the periodic arrangement of cones and $V_{tc}=(\pi
h/3)\left(a^2+ab+b^2\right)$ is the volume of the truncated cone)

\begin{equation}
F = \frac{\pi}{6}\frac{a^2 + ab + b^2}{a\sqrt{b\left(2a +
b\right)}}\;.
\end{equation}

In this regime, the only effective mirror symmetry is $x \rightarrow
-x$, since $\Delta y = h$ or $\Delta y = 0$.

In the limiting case of a cylinder ($a=b$), we correctly get

\begin{equation}
r_1 = r_2 = a\:;\;\; \Delta y_1 = \Delta y_2 = 0\:;\;\; \Delta x_1
= \Delta x_2 = a\:;\;\; \Delta z = \sqrt{3} a\:;\;\; F =
\frac{\pi}{2 \sqrt{3}} = 0.9069\dots
\end{equation}

\vspace{1.0cm}

\subsubsection{Cone-like case}

\vspace{1.0cm}

If $c<c^*$, we are close to the nontruncated cone $c=0$ case.

The two solutions are

\begin{eqnarray}
& & r_1 = \frac{a \left(a + b\right)}{3a + b}\:;\quad \Delta x_1 = 0\\
& & r_2 = \frac{\left(a + b\right)\left(2a + b\right)}{3a + b}\:;\quad
\Delta x_2 = a + b
\end{eqnarray}

yielding

\begin{equation}
\Delta z = \frac{2a \left(2a + b\right)}{3a + b}\;, \quad F =
\frac{\pi}{6}\frac{\left(3a + b\right)\left(a^2 + ab +
b^2\right)}{a\left(a + b\right)\left(2a + b\right)} \;.
\end{equation}

In this regime the only effective mirror symmetry is $y \rightarrow
-y$, since $\Delta x = a + b$ or $\Delta x = 0$ (see Figs. 8 and 11).

In the limiting case of a nontruncated cone ($b=0$), we correctly get

\begin{eqnarray}
& & r_1 = a/3\;, r_2 = 2a/3\:;\quad \Delta x_1 = 0\;, \Delta x_2 =
a\:;\\
& & \Delta y_1 = 2h/3\;, \Delta y_2 = h/3\:;\quad \Delta z =
4a/3\:;\quad F = \frac{\pi}{4} = 0.7854\dots
\end{eqnarray}

and we recover the result in {\it Cone Packing} under {\it Results and
Discussion}.

\vspace{1.0cm}

\subsubsection{Threshold case}

\vspace{1.0cm}

The special value $c=c^*$, i.e., $b = \left(\sqrt{2} - 1\right) a$,
separates the two regimes.

The two solutions are

\begin{eqnarray}
& & r_1 = b\:;\quad \Delta x_1 = 0\:;\quad \Delta y_1 = h \\
& & r_2 = a\:;\quad \Delta x_2 = a + b\:;\quad \Delta y_2 = 0
\end{eqnarray}

yielding

\begin{equation}
\Delta z = \sqrt{2} a\;, \quad F = \frac{\pi}{6} \left(3 - \sqrt{2}\right) = 0.8303\dots\;.
\end{equation}

\end{document}